\documentclass[aps,prl,superscriptaddress,twocolumn,floatfix]{revtex4-1}

\usepackage{graphicx,color}
\usepackage{dcolumn}
\usepackage{bm}
\usepackage{stmaryrd}
\usepackage{latexsym}
\usepackage{amssymb}
\usepackage{amsfonts}
\usepackage{amsmath}
\usepackage{subfigure}
\usepackage{verbatim} %for block comment

\begin{document}
\preprint{APS/123-QED}

\title{The role of interstitial hydrogen in SrCoO$_{2.5}$ antiferromagnetic insulator}

\author{Li Liang}
\affiliation{Institute for Advanced Study, Tsinghua University, Beijing 100084, China}
\affiliation{State Key Laboratory of Low Dimensional Quantum Physics and Department of Physics, Tsinghua University, Beijing 100084, China}
\affiliation{Institute of Electronic Engineering, China Academy of Engineering Physics, Mianyang 621999, China}

\author{Shuang Qiao}
\affiliation{Institute for Advanced Study, Tsinghua University, Beijing 100084,  China}
\affiliation{State Key Laboratory of Low Dimensional Quantum Physics and Department of Physics, Tsinghua University, Beijing 100084,  China}

\author{Shunhong Zhang}
\email{zhangshunhong@tsinghua.edu.cn}
\affiliation{Institute for Advanced Study, Tsinghua University, Beijing 100084,  China}

\author{Jian Wu}
\email{wu@phys.tsinghua.edu.cn}
\affiliation{State Key Laboratory of Low Dimensional Quantum Physics, Department of Physics, Tsinghua University, Beijing 100084,  China}
\affiliation{Collaborative Innovation Center of Quantum Matter, Beijing 100084,  China}

\author{Zheng Liu}
\email{zheng-liu@tsinghua.edu.cn}
\affiliation{Institute for Advanced Study, Tsinghua University, Beijing 100084,  China}
\affiliation{Collaborative Innovation Center of Quantum Matter, Beijing 100084, China}

\date{\today}

\begin{abstract}
Hydrogen exhibits qualitatively different charge states depending on the host material, as nicely explained by the state-of-the-art impurity-state calculation. Motivated by a recent experiment  [Nature 546, 124 (2017)], we show that the complex oxide SrCoO$_{2.5}$ represents an interesting example, in which the interstitial H appears as a deep-level center according to the commonly-used transition level calculation, but no bound electron can be found around the impurity. Via a combination of charge difference analysis, density of states projection and constraint magnetization calculation, it turns out that the H-doped electron is spontaneously trapped by a nonunique Co ion and is fully spin-polarized by the onsite Hund's rule coupling. Consequently, the doped system remains insulating, whereas the antiferromagnetic exchange is slightly perturbed locally.
\end{abstract}

\pacs{Valid PACS appear here}

\maketitle

\begin{figure}
\centering
\includegraphics[height=8cm]{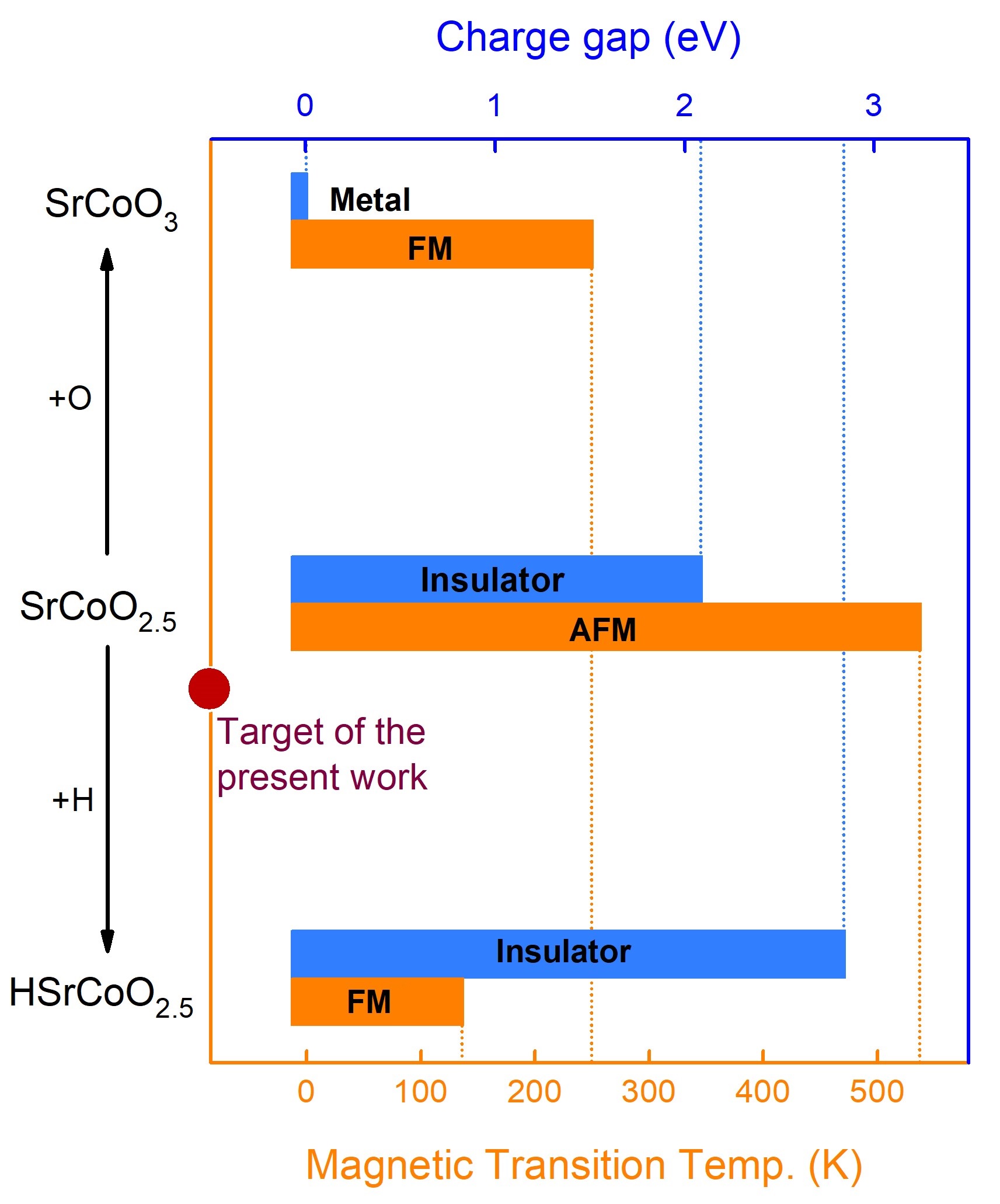}
\caption{A schematic summary of the tri-state phase transitions of SrCoO$_{2.5}$ upon oxidation and hydrogenation as demonstrated in Ref. \cite{Nature17ILG}.}
\label{fig-1}
\end{figure}

The brownmillerite SrCoO$_{2.5}$ is known to be a high-temperature antiferromagnetic (AFM) insulator (T$_N >$ 500 K), reducing from the perovskite SrCoO$_{3}$ ferromagnetic (FM) metal (T$_c \sim$ 250 K) via a long-range ordering of 0.5 oxygen vacancies per formula unit~\cite{JACS06O_intercalate, PRB08neutron,PRL13reversal}. The ordered oxygen vacancies form hollow channels, in which interstitial ions can diffuse with a high mobility~\cite{JCP14Oxy_diffusion}. Recent ionic liquid gating experiment has demonstrated reversible insertion and extraction of interstitial hydrogen ions into the hollow channels, reaching a new HSrCoO$_{2.5}$ insulating phase that exhibits magnetic hysteresis below 125 K~\cite{Nature17ILG}. A schematic summary of these rich electronic and magnetic transitions is shown in FIG.~\ref{fig-1}, which presents SrCoO$_{2.5}$ as a useful platform for applications such as magnetoelectric devices and solid-state fuel cells.

Despite a number of studies on SrCoO$_{2.5+\delta}$ ($0\leq\delta\leq 0.5$)~\cite{JACS06O_intercalate,NMat13SrCoOx,AM13phase_transition}, the physics of hydrogenated SrCoO$_{2.5}$ remains largely unexplored, because this new phase was not accessed before via traditional growth methods. In contrast to the oxidation side, where an insulator-to-metal transition occurs, the insulating gap of SrCoO$_{2.5}$ appears intact upon hydrogenation. More peculiarly, according to the magnetization and soft X-ray magnetic circular dichroism measurements, HSrCoO$_{2.5}$ is a weak FM insulator~\cite{Nature17ILG}, which is rather rare in nature.

Here, we aim to present a theoretical understanding of the role of interstitial hydrogen in SrCoO$_{2.5}$. Based on first-principles calculations, we show that the inserted hydrogen acts as an electron donor, but the doped electron is not mobile. Instead, it is spontaneously trapped around a Co site, reducing the local Co valence state from +3 to +2. Due to the multivalent nature of Co ions, this Co$^{2+}$ state is stable. The consequences are: (a) no impurity state is introduced around the band edge; and (b) the AFM order is slightly perturbed locally. These results not only put forth a theoretical basis to understand the experimental observations on HSrCoO$_{2.5}$, but also shed new light on the impurity calculation methodology when strongly-correlated electrons are involved.

First-principles calculations are performed within the framework of density functional theory (DFT) as implemented in the Vienna \textit{Ab initio} Simulation Package (VASP)~\cite{PRB94VASP}.  We employ projector augmented wave method~\cite{PRB94PAW} to treat the core electrons, and the generalized gradient approximation as parametrized by Perdew, Burke, and Ernzerhof (PBE) \cite{PRL96GGA-PBE} for the exchange-correlation functional of the valence electrons. A kinetic energy cutoff of 500 eV is found to achieve numerical convergence. The strong on-site Coulomb repulsion of Co-3$d$ orbitals are treated by the simplified DFT+U method~\cite{PRB1998LDA+U}. An effective U-J value of 6.5 eV is used for all calculations following Ref.~\cite{JPhys14DFT+U}. The initial magnetic moments of any two neighboring Co ions are set anti-parallel, which reproduces the so-called G-type AFM configuration as determined by previous neutron scattering measurements~\cite{PRB08neutron}. The spin density is then treated self-consistently in the DFT calculation. The self-consistent electronic iteration are performed until the total energy change is smaller than 10$^{-5}$ eV.

\begin{figure}
\centering
\includegraphics[height=8cm]{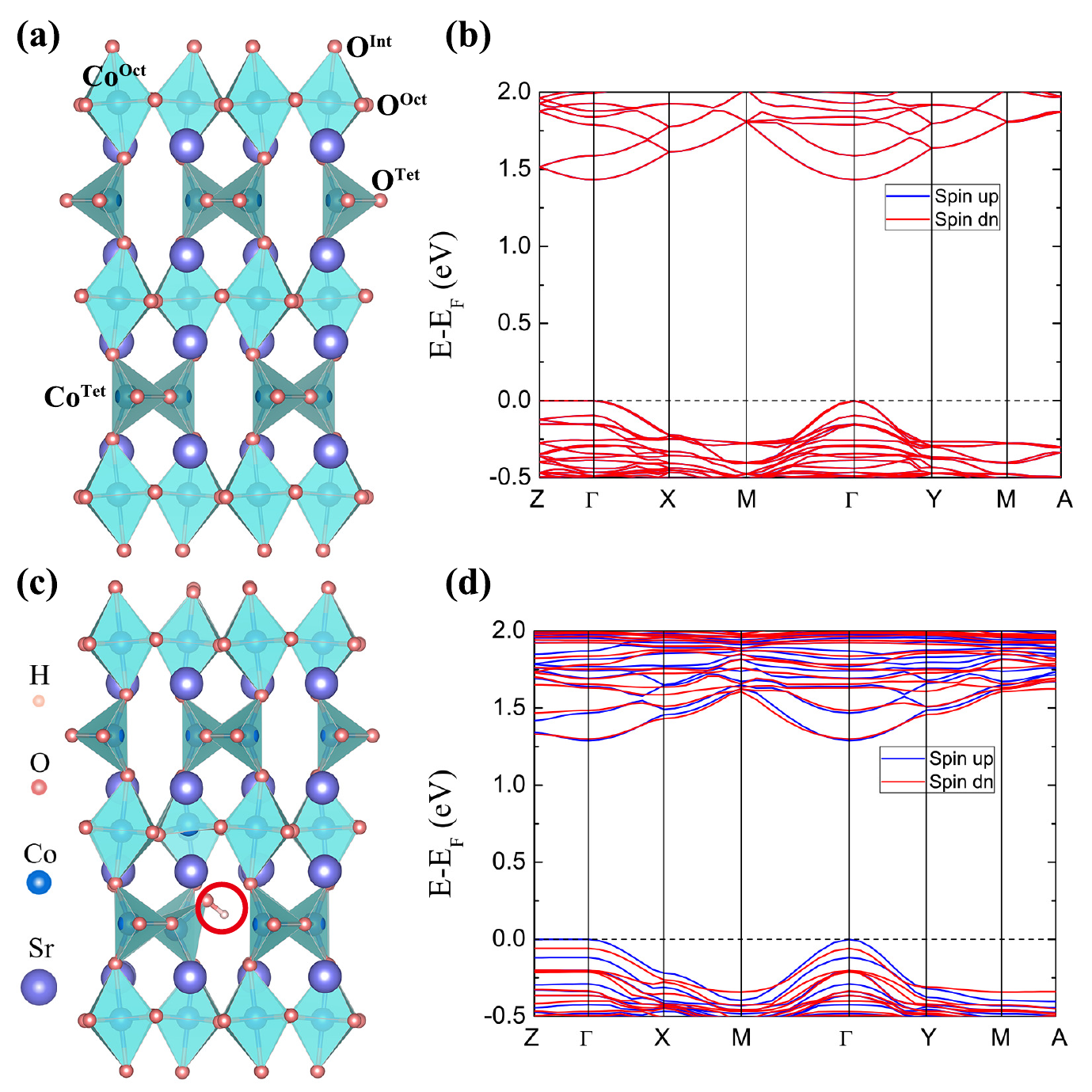}
\caption{Crystal structures of SrCoO$_{2.5}$ (a) before and (b) after H doping. Red circle marks the lowest-energy position of the interstitial H locating in the hollow channel. (c) and (d), the corresponding DFT+U band structures.}
\label{fig-2}
\end{figure}

The crystal structure of SrCoO$_{2.5}$ is shown in FIG.~\ref{fig-2}(a). There are two inequivalent Co sites, one is in a tetrahedral CoO$_4$ coordination (denoted as Co$^\texttt{Tet}$) and the other is in an octahedral CoO$_6$ coordination (denoted as Co$^{\texttt{Oct}}$). Accordingly, the O atoms can be classified into three types: O in the tetrahedral plane (O$^{\texttt{Tet}}$), O in the octahedral plane (O$^{\texttt{Oct}}$), and O in the inter-layer space (O$^{\text{Int}}$). Both the lattice and atomic positions are fully relaxed without subjecting to a specific symmetry group, until the forces are smaller than 10$^{-2}$ eV/\AA. The optimized structure is found to fit best to the Ima2 space group.  %The optimized structure of SrCoO$_{2.5}$ is appended as a supplementary material of this paper.

We note that some ambiguity remains in the experimentally refined structure data, primarily implying slight differences in the arrangement of the CoO$_4$ tetrahedra~\cite{PRB08neutron, JAP15Tetra}. We do not intend to address this controversy here, considering that it is a minor effect compared to the H-induced distortion of the CoO$_4$ tetrahedra [FIG.~\ref{fig-2}(c)].

To simulate hydrogen doping, we construct a supercell containing 32 Sr, 80 O and 32 Co atoms, and introduce one H within, corresponding to a low H-concentration case (marked by the red dot in FIG.~\ref{fig-1}). The distance between two nearest H atoms under the periodic boundary condition is larger than 10 \AA\ and a single $\Gamma$ point is used to sample the mini Brillouin zone. The atomic positions in the supercell are relaxed again, while the lattice constants are fixed to the primitive-cell optimized values.

We first determine the most stable H position. Previous study suggests that the ions under liquid gating are preferably inserted into the hollow channels~\cite{Nature17ILG}. We have tested several different initial positions, and the lowest total energy is obtained when the H binds with an O$^{\texttt{Int}}$ [FIG.~\ref{fig-2}(c)]. In comparison, the total energy is 0.19 eV higher when the H binds with an O$^{\texttt{Oct}}$, and 0.45 eV higher when the H binds with an O$^{\texttt{Tet}}$.  Our structural relaxation does not find any local energy minimum for the H to bind with a Co or Sr. The formation of an H-O$^{\texttt{Int}}$ bond is found to induce distortion of the associated CoO$_4$  tetrahedron and weaken the associated O$^{\texttt{Int}}$-Co$^{\texttt{Oct}}$ bond.

We then calculate the electronic band structures before and after H doping. Figure~\ref{fig-2}(b) clearly reveals a band gap of the size $\sim$ 1.4 eV of the pristine SrCoO$_{2.5}$. This gap value is larger than the previous DFT+U results, which varied the onsite Coulomb repulsion U from 3 to 5 eV~\cite{PRB08neutron}, yet smaller than the experimental value $\sim$ 2 eV determined optically~\cite{Nature17ILG}.  Considering that the Co ion nominally has a +3 valence state, leaving 6 d-electrons per Co site, an insulating ground state naturally fits in the band theory. The spin-up and spin-down bands are completely degenerate, as expected for an ideal AFM spin density, which is invariant under the combination of time reversal and sublattice inversion.  The magnetic moments of both Co$^{\texttt{Oct}}$ and Co$^\texttt{Tet}$ converge to $\sim 3.1\ \mu_B$, slightly larger than the previous calculation results obtained with a smaller U~\cite{SR17SrCoO3}. The previous neutron scattering data showed temperature dependence and a small difference of the magnetic moment values for Co$^{\texttt{Oct}}$ and  Co$^\texttt{Tet}$ - at T = 10K, the values are 3.12 $\mu_B$ and 2.88 $\mu_B$, respectively~\cite{PRB08neutron}. These values roughly coincide with a Co$^{3+}$ S = 2 high-spin state with some hybridization with the O p-orbitals~\cite{PRB08neutron}.

It is surprising to find that after H doping, the band-edge properties barely change [FIG.~\ref{fig-2}(d)]. In particular, we do not observe any impurity band within the gap or around the band edge, and the Fermi level remains at the valence band maximum. The most noticeable effect is a splitting of the spin degeneracy, signaling the breaking of the AFM symmetry. A quick check of the spin density shows that the supercell carries a net magnetic moment of $\sim$ 0.91 $\mu_B$, close to the contribution of a single electron.

A widely-used calculation method~\cite{PRB01defect,Nature03H} to characterize the electrical activity of impurities in conventional semiconductors or  insulators is to purposely change the electron number in the supercell and restore the charge neutral condition with a homogeneous charge background. Such a simulation can be regarded as artificially liberating some electrons or holes into a free carrier state, leaving a charged impurity state behind [e.g. a H$^+$  (H$^-$) impurity state by removing (adding) one electron]. Depending on the choice of the Fermi level that accommdotates the liberated electrons or holes, the energy difference between such a constraint electronic system and the original fully-relaxed one defines an estimated activation energy for the system becoming conductive, which is commonly termed as the defect transition level~\cite{PRB01defect,Nature03H}. Following this recipe, we find that the H$^0$/H$^+$ transition energy is 0.95 eV below the conduction band minimum and the H$^0$/H$^-$ transition energy is 0.52 eV above the valence band maximum, both sufficiently deep within the gap. Although these values are subject to various uncertainties, e.g. band gap errors and long-range interactions between the charged impurities, it appears at the first sight that the interstitial H can be reasonably classified as a deep-level impurity, and thus the preserving of an insulating state under hydrogenation becomes natural. 

However, a second thought alerts that a clean band gap after H-doping is distinct from what a deep-level impurity typically manifests. For example, interstitial H is known to be a deep-level impurity in MgO~\cite{APL02MgO_H}. Accordingly, a bound state deep inside the band gap presents, which reflects the fact that the electron is trapped by the highly localized impurity potential and thus is difficult to get activated into a free carrier. In our case, such an in-gap impurity level is always absent no matter whether the impurity is charged or not (c.f. FIG.~\ref{fig-4}). One might contend that the impurity level could lie deep inside the valence or conduction bands, but the difficulty is to satisfy the electron counting. Recall that the interstitial H introduces one extra electron. To keep the Fermi level within the gap, the only possibility is that the spin degeneracy of the impurity level is somehow removed, leading to one spin-polarized level inside the valence bands and the other inside the conduction bands. In this way, the emergence of a net magnetic moment close to 1 Bohr magneton is clarified as well.

\begin{figure}
\centering
\includegraphics[height=8cm]{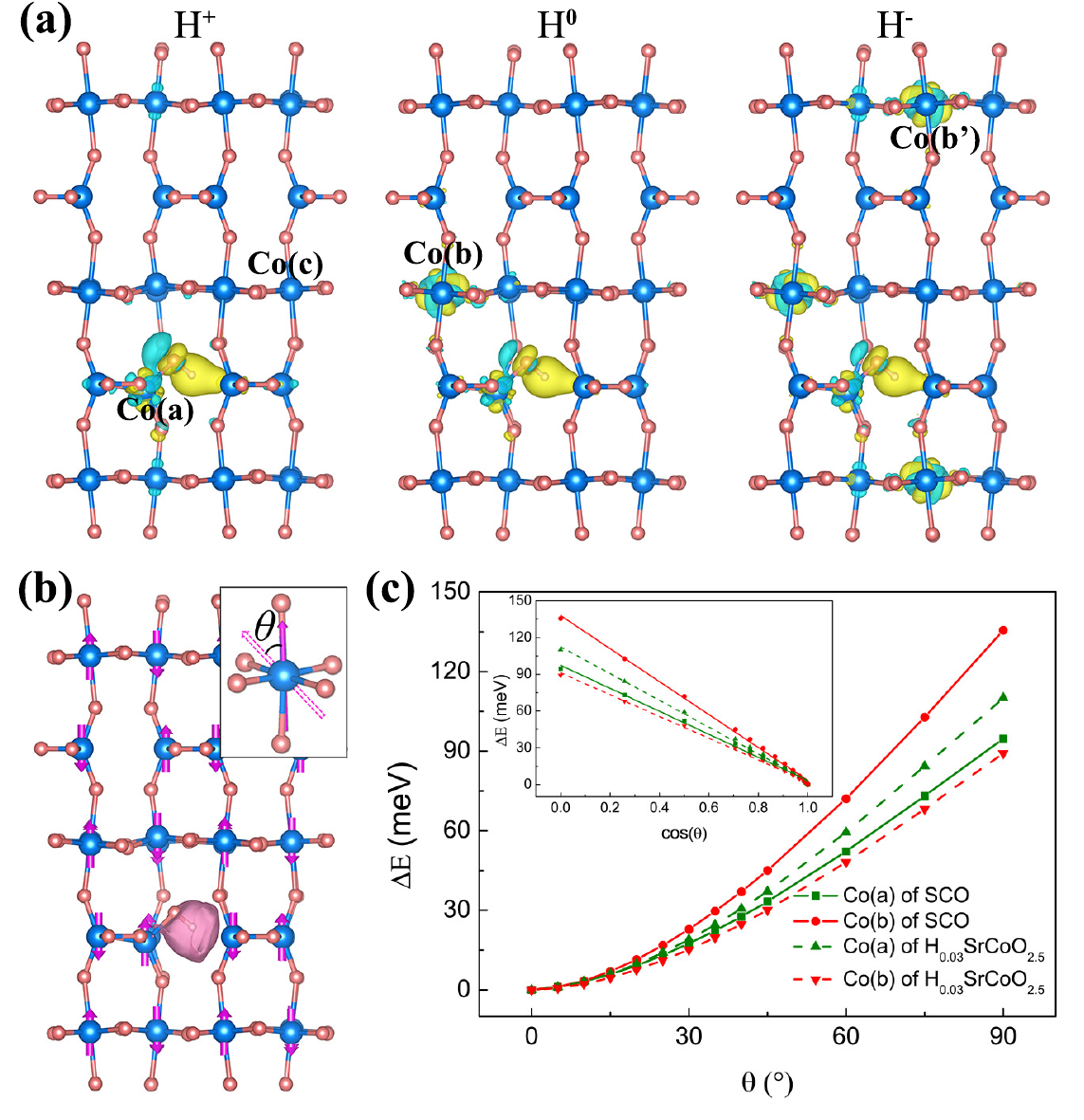}
\caption{(a) Charge difference induced by an interstitial H in different charge states. (b) The electrostatic potential induced by the H$^+$ ion. The isovalue contour indicates the range, within which the potential has decayed by one order of magnitude from the maximum value. Also shown is the ground-state AFM configuration of the Co ions. The spin rotation angle $\theta$ of a single Co ion is defined in the inset. (c) Change of total energy $\Delta E$ by rotating different Co sites before and after H doping. The inset shows the $\Delta E$-$\cos \theta$ fitting}.
\label{fig-3}
\end{figure}

\begin{figure}
\centering
\includegraphics[height=6cm]{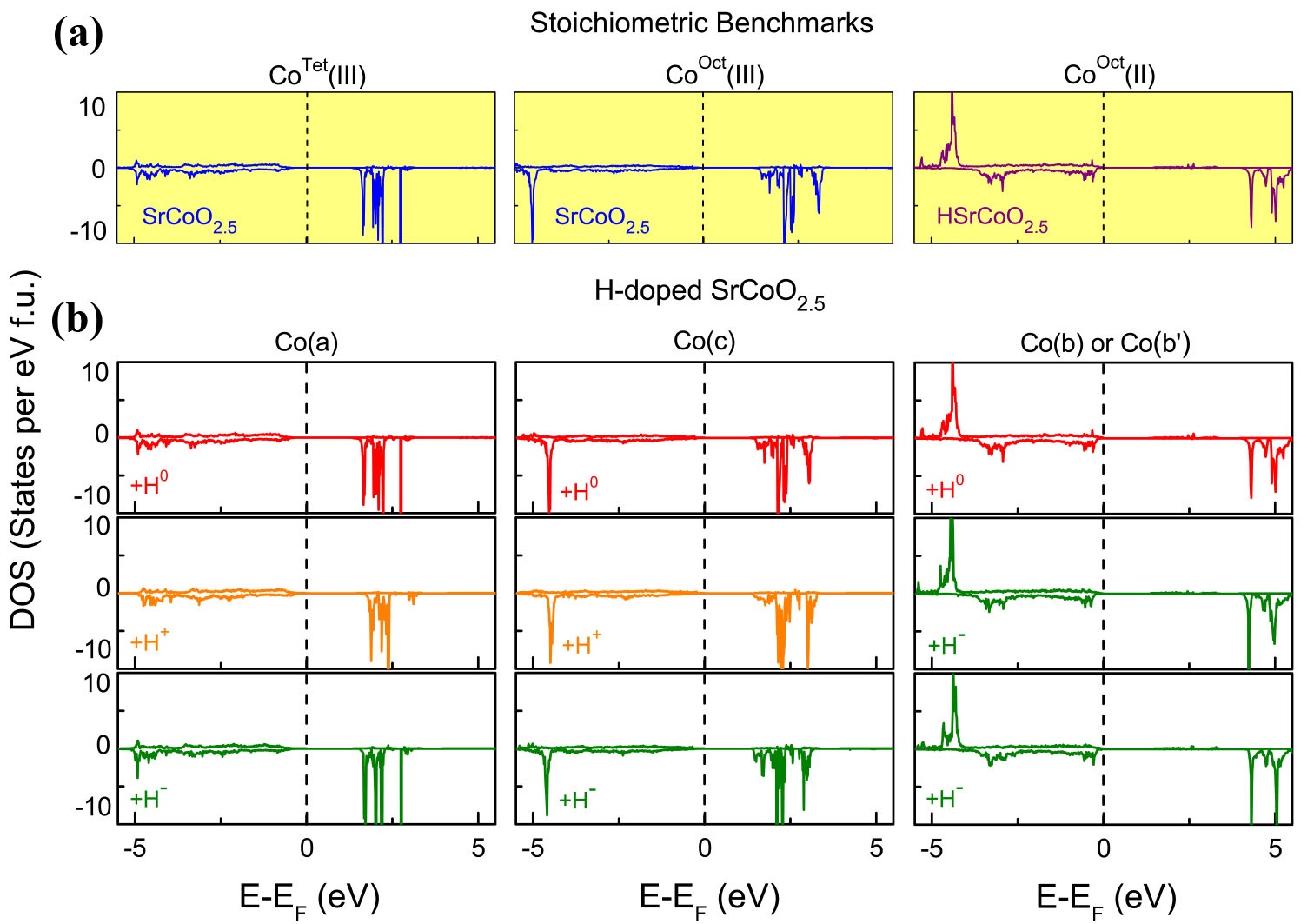}
\caption{Comparison of DOS projected on selective Co sites in (a) stoichiometric SrCoO$_{2.5}$ and HSrCoO$_{2.5}$; and (b) the H-doped supercell as shown in FIG.~\ref{fig-3}.}
\label{fig-4}
\end{figure}

Then, how could such a large spin splitting at least of the size of the band gap occur? This puzzle is resolved by noticing that the doped electron is trapped by a Co ion instead of the interstitial H. Consequently, the strong Hund's rule coupling of the 3d orbitals polarize the electron spin. In specific, given that the 6 d-electrons of a Co$^{3+}$ form a high-spin state, the spin of the trapped electron is enforced to lie in antiparallel, effectively giving rise to a Co$^{2+}$ S = 3/2 state. Figure~\ref{fig-3}(a) plots the charge difference due to the introduction of an interstitial H with the charge state H$^+$, H$^0$ and H$^-$, respectively. For the H$^+$ case, only one proton is introduced and the electron number does not change. We observe a charge redistribution around the H, the O$^\texttt{int}$ bonded to the H, and the Co$^\texttt{Tet}$ bonded to the hydroxyl [labeled as Co(a) in FIG.~\ref{fig-3}(a)], which is attributed to the perturbation of the proton potential. The other Co sites [e.g. Co(c) labeled in FIG.~\ref{fig-3}(a)] are almost unperturbed. For the H$^0$ case, the same charge redistribution around the impurity also presents. In addition, one Co$^\texttt{Oct}$ away from the H [labeled as Co(b) in FIG.~\ref{fig-3}(a)] possesses extra electron density, which is attributed to the H-doped electron. For the H$^-$ case, one more Co [labeled as Co(b') in FIG.~\ref{fig-3}(a)] possesses extra electron density.

It is intriguing that the extra electrons are bound to neither the H nor the nearest-neighbor Co, but some distant Co ions. Figure~\ref{fig-3}(b) plots the electrostatic potential difference due to the introduction of an interstitial H$^+$, i.e. the proton potential. It is clear that Co(b) and Co(b') are beyond the impurity potential range. We do not see any direct connection between the locations of Co(b), Co(b') and H. We have observed that the doped electron can localize around another Co site when the initial conditions of the iteration slightly differ. Thus, the doped Co site picked by the self-consistent iterations may subtly depend on some finite-size effects of our supercell. In the thermodynamic limit, we consider that this is a spontaneous symmetry breaking process - the doped electron falls into a local energy minimum randomly. From this perspective, interstitial H in SrCoO$_{2.5}$ is \emph{not} a deep-level impurity, but rather a good $n$-type donor, given that it has effectively donated its electron into the host lattice. It is the multivalent ``goblin'' - the original meaning of cobalt in German - that traps the doped carrier, giving rise to the large activation energy.

We can further confirm from the atomic projected density of states (PDOS) that the doped electron is absorbed into a Co$^{3+}$ forming a Co$^{2+}$, while the other Co$^{3+} $ ions are not affected. We first reproduce the PDOS of Co in stoichiometric SrCoO$_{2.5}$ and HSrCoO$_{2.5}$ \cite{Nature17ILG} as the benchmark of Co$^{3+}$ and Co$^{2+}$ in the brownmillerite environment [FIG.~\ref{fig-4}(a)]. They exhibit distinct features - especially, the gap size shown in the Co$^{2+}$ PDOS is much larger than in the Co$^{3+}$ PDOS. An increase of the insulating gap in HSrCoO$_{2.5}$ was indeed observed in experiment~\cite{Nature17ILG}. In our H-doped supercell, the PDOS of Co(b) and Co(b') closely resembles Co$^{2+}$ [FIG.~\ref{fig-4}(b), right column]. All the other Co atoms, including Co(a), display features similar to Co$^{3+}$ [FIG.~\ref{fig-4}(b), left and central columns]. 

The remaining question is whether H doping frustrates the AFM structure. It is worth mentioning that doping an AFM insulator usually spawns various exotic electronic phases owing to the competition between the electron kinetic energy and the AFM exchange energy, as long studied in the context of high-temperature cuprate superconductors. If so, the magnetic order of  H$_x$SrCoO$_{2.5}$ would become rather complicated, depending on the H concentration. We analyze this problem by manually rotating the spin orientation of Co(a) and Co(b) starting from the ground state AFM configuration in a series of constraint noncollinear magnetization calculations [inset of FIG.~\ref{fig-3}(b)]. For the pristine SrCoO$_{2.5}$, there is no doubt that the total energy change ($\Delta E$) as a function of the rotation angle ($\theta$) reflects the overall exchange strength of the Co$^\texttt{Tet}$ and Co$^\texttt{Oct}$ sites with their neighbors. Figure~\ref{fig-3}(c) shows the calculated data points. The energy is the lowest at $\theta$ = 0, and increases monotonously when the configuration deviates from the AFM ground state. A good linear relation between $\Delta E$ and cos$\theta$ is found, indicating a primary Heisenberg-type exchange. We note that the ratio between $\Delta E [Co^\texttt{Tet}]$ and $\Delta E[Co^\texttt{Oct}]$ is roughly 2/3, because of the CoO$_4$ and CoO$_6$ coordination. Both the previous paper~\cite{JPhys14DFT+U} and our own calculation indicate that the Co-O-Co exchange is very isotropic. We then apply the same calculation to the doped supercell. The data indicates that the linear fitting still works, with only quantitative change of the $\Delta E$-cos$\theta$ slope. The overall AFM coupling between Co(a) and its neighbors slightly increases, possibly due to the distortion of the bonding geometry. The overall AFM coupling between Co(b) and its neighbors is reduced more remarkably, to a value close to  $Co^\texttt{Tet}$. If we formulate the magnetic energy as $\Delta E=\textbf{h}^{MF}\cdot\textbf{S}_{Co(b)}$, in which $\textbf{h}^{MF}$ is the molecular field acted on Co(b), the reduction of the Co(b) magnetic moment from 3d$^6$ S = 2 to 3d$^7$ S = 3/2 accounts for a large fraction of the change, whereas  $\textbf{h}^{MF}$, i.e. the surounding magnetic order and the AFM exchange strength, is only slightly weakened.

In conclusion,  we show that interstitial H in the AFM insulator SrCoO$_{2.5}$ is an electron donor, but the kinetic energy of the doped electron is quenched by the multivalent Co ion. In this view, the robust insulating gap observed in the H$_{1-\delta}$SrCoO$_{2.5}$ sample can be understood. Also, as long as $\delta \neq 0$ or 1, the difference between the Co$^3+$ and Co$^2+$ magnetic moment will give rise to some FM signals. Nevertheless, we should point out that the present calculation cannot provide a quantitative explanation on  the $\sim$125 K magnetic transition experimentally observed in H$_{1-\delta}$SrCoO$_{2.5}$ ~\cite{Nature17ILG}, which is only 1/5 of the Neel temperature of SrCoO$_{2.5}$. It remains an open question whether additional complexities emerge when the H concentration increases, i.e. the correlation between the doped electrons becomes important. Leaving aside the intermediate doping region that is difficult for first-principles simulation, some insight could be obtained by applying our approach to the H-rich limit, studying the role of a single H vacancy in HSrCoO$_{2.5}$. To carry out this calculation, further experimental input about the refined atomic and magnetic structures of HSrCoO$_{2.5}$ is demanded. 

\section*{Acknowledgements}
We would like to thank J.-W. Mei, Z.-Y. Weng, M. Coey, J.Y. Zhu and P. Yu for helpful discussion. This work is supported by Tsinghua University Initiative Scientific Research Program and NSFC under Grant No. 11774196. L.L. acknowledges support from NSFC (Grant No. 11505162). S.Z. is supported by the National Postdoctoral Program for Innovative Talents of China (BX201600091) and the Funding from China Postdoctoral Science Foundation (2017M610858).

\end{document}